\documentclass[review]{elsarticle}

\usepackage{hyperref}

\usepackage{amsmath}
\usepackage{amssymb}
\usepackage{bm}
\usepackage{multirow}
\usepackage{textcomp}

\journal{Computer Physics Communications}

\DeclareMathOperator*{\minmod}{minmod}

\newcommand{\NN}{\mathbb{N}}
\newcommand{\RR}{\mathbb{R}}

\begin{document}

\begin{frontmatter}

\title{An accurate scheme to solve cluster dynamics equations using a Fokker-Planck approach}

\author[cea]{T.~Jourdan\corref{cor1}}
\ead{thomas.jourdan@cea.fr}

\author[cermics,matherials]{G.~Stoltz}
\author[navier,matherials]{F.~Legoll}
\author[cermics]{L.~Monasse}

\cortext[cor1]{Corresponding author. Tel.:+33 1 69 08 73 44. Fax: +33 1 69 08 68 67}

\address[cea]{DEN-Service de Recherches de M\'etallurgie Physique, CEA, Universit\'e Paris-Saclay, F-91191 Gif-sur-Yvette, France}
\address[cermics]{Universit\'e Paris-Est, Cermics (ENPC), F-77455 Marne-la-Vall\'ee Cedex 2, France}
\address[navier]{Laboratoire Navier, Ecole des Ponts ParisTech, F-77455 Marne-la-Vall\'ee Cedex 2, France}
\address[matherials]{MATHERIALS project-team, INRIA Paris, 2 rue Simone Iff, CS 42112, 75589 Paris Cedex 12, France}

\begin{abstract}
    
  We present a numerical method to accurately simulate particle size
  distributions within the formalism of rate equation cluster
  dynamics. This method is based on a discretization of the associated
  Fokker-Planck equation. We show that particular care has to be taken
  to discretize the advection part of the Fokker-Planck equation, in
  order to avoid distortions of the distribution due to numerical
  diffusion. For this purpose we use the Kurganov-Noelle-Petrova
  scheme coupled with the monotonicity-preserving reconstruction MP5,
  which leads to very accurate results. The interest of the method is
  highlighted on the case of loop coarsening in aluminum. We show that
  the choice of the models to describe the energetics of loops does
  not significantly change the normalized loop distribution, while the
  choice of the models for the absorption coefficients seems to have a
  significant impact on it.

\end{abstract}


\end{frontmatter}


\section{Introduction}
\label{sec:introduction}

Nucleation, growth and coarsening of particles are three mechanisms
which appear in gases, liquids and solids. For example, in solids the
formation and evolution of secondary-phase precipitates or point
defect clusters such as voids, dislocation loops and stacking fault
tetrahedra can be described by a combination of these three
mechanims. The evolution of a system in terms of particle size
distributions thus can be modeled by coupling methods specifically
designed to describe one particular
mechanism~\cite{Langer1980,Maugis2005}. An alternative modeling approach
relies on the observation that the three processes result from the
absorption and emission of mobile particles by other particles (birth
and death process). To avoid the approximations which are needed when
methods are coupled, it is therefore interesting to directly model the
evolution of particles due to these absorption and emission processes.

Only a few methods allow a seamless description of the nucleation,
growth and coarsening processes by simulating the absorption and
emission of particles over time scales comparable to
experiments. Atomistic and Object kinetic Monte Carlo (kMC) methods
are classical methods to perform such
tasks~\cite{soisson_cu-precipitation_2007,Domain2004c,Jourdan2010a}. Their
computational cost however limits the size of the systems which can be
simulated, so that it is impossible to simulate the whole evolution of
a realistic microstructure. Cluster dynamics, or rate equation cluster
dynamics (RECD), appears to be an efficient alternative which can
accurately describe a locally homogeneous system with a moderate
volume fraction of particles. In RECD, the system is modeled as a gas
of particles (often called clusters), which can migrate in the solvent
(the matrix in case of a solid), associate and dissociate~\cite{ortiz2007_2,clouet_modeling_2009}. Comparisons
with kMC methods have shown that, provided the parameters of the RECD
models are carefully chosen, particle size distributions (also called
cluster distributions in the following) obtained by the two methods
are very similar~\cite{Jourdan2010,Jourdan2012}.

RECD models are written as a set of ordinary differential equations
(ODEs) which correspond to the birth and death process. The variables
are the concentrations of each cluster type (see details below).
Despite their simplicity, RECD equations are notably difficult to
solve. They are in general very stiff, so that the integration in time
should be performed with implicit methods~\cite{Hairer1996}. These
methods require solving linear systems based on the Jacobian matrix of
the ODE set. Since there is one equation per cluster size, the number
of equations and thus the size of the Jacobian matrix can be larger
than $10^6$ to handle clusters of a few tens of nanometers. This can
become prohibitively large in terms of computational time and memory
use, even when the sparsity of the matrix is taken into
account~\cite{Jourdan2014}. Alternatively, stochastic methods can be
used~\cite{Marian2011,Gherardi2012,Dunn2015}: they do not require the
storage of the Jacobian matrix but the time step is often much smaller
than with deterministic methods.

In order to reduce the number of equations to solve, several methods
have been introduced. They can be divided into two main classes, the
``grouping'' methods and the methods based on a partial differential
equation (PDE) obtained in the continuum limit. The former approach,
first developed by Kiritani~\cite{Kiritani1973}, consists in solving
equations for the moments of the distribution over groups of
equations. It was shown by Golubov \emph{et
  al.}~\cite{Golubov2001,Ovcharenko2003} that using at least the
zeroth and first moments of the distribution was necessary to obtain
accurate results. The second class of methods is based on the
observation that, for large cluster sizes, the master equation can be
approximated by a Fokker-Planck equation. By discretizing this
equation on a sufficiently coarse mesh, a significant reduction in the
number of equations can be achieved. However, assessment of the
accuracy of this class of methods remains scarce compared to grouping
methods. It is sometimes even argued that Fokker-Planck based methods
inherently cannot be accurate~\cite{Golubov2012}.

\medskip

In this article we show that, by carefully discretizing the
Fokker-Planck equation, it is possible to achieve high accuracy in
the simulation of cluster distributions. A proper spatial discretization of the advection term turns out to be critical to obtain qualitatively and quantitatively accurate results. As an application, we
simulate the coarsening of loops, for which an analytical form of the
loop distribution is known under some approximations. These various
approximations are discussed in light of our RECD calculations.

Our article is organized as follows. We recall the RECD model and its
continuum Fokker-Planck limit in
Section~\ref{sec:master-equation-FP}. Various discretization schemes
for the Fokker-Planck equation are discussed in
Section~\ref{sec:num-scheme-FP}, with an emphasis on the spatial
discretization of the advection term. Numerical results are then
presented in Section~\ref{sec:results-and-discussion}. First, we
compare in Section~\ref{sec:cluster-quenched-vacancies} the various
numerical schemes on a simple test case where reference simulations
can be performed, in order to validate the scheme presented in
Section~\ref{sec:mp5}. We next use this scheme in
Section~\ref{sec:loop-coarsening} to assess the relevance of various
modeling assumptions to describe loop coarsening in aluminum. Our
conclusions are collected in Section~\ref{sec:conclusion}.

\section{Master equation and the Fokker-Planck approximation}
\label{sec:master-equation-FP}

\subsection{Master equation approach to RECD}

For the sake of simplicity, we consider in the following the evolution
of cluster concentrations due to the absorption and emission of
monomers. Clusters are assumed to contain only one type of
species. For secondary-phase precipitates or point defect clusters in
solids, this species represents a solute, a vacancy or a
self-interstitial. Clusters thus can be referred to by a single index $n =
1, \dots, N$ and their concentration is denoted by $C_n$. More general
cases (clusters with multiple species and absorption of mobile
clusters) are described in~\cite{Jourdan2014}. In the case considered
here, the RECD equations to solve are
\begin{align}
  \label{eq-recd-nsup2}
  \frac{\mathrm{d}C_n}{\mathrm{d}t} &= \beta_{n-1}C_{n-1}C_1 - \alpha_n C_n - \beta_n C_n C_1 + \alpha_{n+1}C_{n+1} \qquad \text{for any $n \geq 2$},\\
  \label{eq-recd-neq1}
  \frac{\mathrm{d}C_1}{\mathrm{d}t} &= -2\beta_1 C_1 C_1 + 2 \alpha_2 C_2 - \sum_{n \geq 2} \left[\beta_{n} C_n C_1 - \alpha_{n+1}C_{n+1} \right].
\end{align}
The coefficients $\beta_n$ and $\alpha_n$ represent the absorption and emission
rates of monomers respectively. For example, if clusters are spherical
and if the reactions between clusters are controlled by diffusion, these
coefficients read~\cite{waite_theoretical_1957,Russel1971}
\begin{align}
  \beta_n &= 4\pi r_n D_1, \label{eq-betan}\\
  \alpha_n &= \frac{\beta_{n-1}}{V_{\mathrm{at}}} \exp{\left(-\frac{F_{n}^b}{kT}\right)} , \label{eq-alphan}
\end{align}
where $r_n$ is a reaction distance, $V_{\mathrm{at}}$ is the atomic volume,
$F_n^b$ is the binding free energy of a monomer to a cluster containing
$n-1$ monomers, $k$ is the Boltzmann constant, $T$ is the
temperature and $D_1$ is the diffusion coefficient
of the monomer given by
\begin{equation}
  \label{eq-diffusion-coefficient-monomer}
 D_1 = D_{1,0} \exp{(-E^{\mathrm{m}}_1/kT)}. 
\end{equation}
In this expression, $D_{1,0}$ is a diffusion prefactor and $E^{\mathrm{m}}_1$ is the migration
energy of the monomer.

Using~\eqref{eq-recd-nsup2}--\eqref{eq-recd-neq1}, it is straightforward to check that the total quantity of matter $S$ is conserved, \emph{ie}
\begin{equation}
  \label{eq-conservation}
  \frac{\mathrm{d} S}{\mathrm{d}t} = \frac{\mathrm{d}}{\mathrm{d}t}\left(\sum_{n\ge 1} n C_n \right) = 0.
\end{equation}

\subsection{Fokker-Planck approach to RECD}

Using a Taylor expansion, it can be shown~\cite{Goodrich1964,Terrierxxxx} that Eq.~\eqref{eq-recd-nsup2} can be approximated, for $n \gg 1$, by a Fokker-Planck equation of the form
\begin{equation}
  \label{eq-fokker-planck}
  \frac{\partial}{\partial t} C(x,t) = -\frac{\partial }{\partial x} \left[F(x,t) C(x,t) \right]+ \frac{\partial^2}{\partial x^2} \left[ D(x,t) C(x,t)\right],
\end{equation}
with $x_n = n$, $C(x_n,t) \approx C_n(t)$ and
\begin{align}
  \label{eq-Fx}
  F(x,t) &=  \beta(x) C_1(t) - \alpha(x),\\
  \label{eq-Dx}
  D(x,t) &= \frac{1}{2} \left[\beta(x) C_1(t) + \alpha(x)\right].
\end{align}
To perform this approximation, it has been assumed that there exist two functions $\alpha$ and $\beta$ such that $\alpha_n = \alpha(x_n)$ and $\beta_n = \beta(x_n)$. In practice this is the case for large values of $n$, for which $F_n^b$ and $r_n$ in Eqs.~\eqref{eq-betan} and~\eqref{eq-alphan} can be expressed analytically as functions of $n$ (an example is given in Section~\ref{sec:loop-coarsening}).
The transformation to a Fokker-Planck equation is interesting as such, since the two physical
ingredients present in the rate equations are clearly highlighted. The
advection term, proportional to $F(x,t)$, is related to cluster
growth. The diffusive part, proportional to $D(x,t)$, is related to
cluster size fluctuations due to random absorption and emission of
monomers. It is responsible for a broadening of the cluster
distribution. If clusters contain multiple species, a multidimensional Fokker-Planck equation can be written in a very similar way~\cite{Sharafat1990}. The method proposed in this work for the advective term can simply be applied componentwise. For the diffusion part, cross-terms of the type $\partial^2 ./ \partial x \partial y$ appear if clusters containing multiple species are mobile. The treatment of such terms is not addressed in the present article.

Since we intend to use a Finite Volume formulation in Section~\ref{sec:num-scheme-FP}, it is convenient to rewrite Eq.~\eqref{eq-fokker-planck} in a conservative form, as a function of the flux $J$. It also permits to exhibit the advection and diffusion fluxes $J^{\mathrm{a}}$ and $J^{\mathrm{d}}$:
\begin{equation}
  \label{eq-FP-with-fluxes}
  \frac{\partial C}{\partial t} = -\frac{\partial J}{\partial x} = -\frac{\partial J^{\mathrm{a}}}{\partial x} -\frac{\partial J^{\mathrm{d}}}{\partial x}
\end{equation}
with
\begin{align}
  \label{eq-Ja}
  J^{\mathrm{a}}(x,t,C(x,t)) &= F(x,t) C(x,t) \\
  \label{eq-Jd}
  J^{\mathrm{d}}(x,t,C(x,t)) &= -\frac{\partial }{\partial x} \left[D(x,t) C(x,t) \right].
\end{align}

In general no analytical solution can be found for the partial
differential equation~\eqref{eq-fokker-planck}, but it can be discretized along $x$ and
solved numerically as an ODE set. The advantage with respect to~\eqref{eq-recd-nsup2}--\eqref{eq-recd-neq1} is that a coarse mesh size $\Delta x$ may often be used. The resulting set of ODEs is thus posed in a space of lower dimension than~\eqref{eq-recd-nsup2}--\eqref{eq-recd-neq1}. Usually, due to the strong stiffness
of the ODE set, implicit methods with backward differentiation formula
or Runge-Kutta time stepping are used~\cite{Hairer1996}.

Discretizing and directly solving the Fokker-Planck equation for
cluster dynamics is not a very common approach. Among the few attempts
that can be found in the literature, one can note the works by
Ghoniem and Sharafat~\cite{Ghoniem1980,Ghoniem1999} and, more
recently, Turkin and Bakai~\cite{Turkin2006}. In these works, the
advective term is taken care of by centered finite differences. This approach has
been shown to give satisfactory results when the advective flux is
small, such as for the simulation of
aging~\cite{Turkin2007,Jourdan2014}. However, for large advection
terms, this scheme is known to be unstable~\cite{Fletcher1991}, which manifests itself through spurious oscillations in the cluster distributions as will be
shown in Sec.~\ref{sec:cluster-quenched-vacancies}. To solve this
problem, upwind schemes~\cite{Fletcher1991,Ozkan2000} or related schemes such as the Chang-Cooper scheme~\cite{Chang1970} can be used~\cite{Rempel2009,Jourdan2014}. Essentially, the Chang-Cooper scheme automatically
and seamlessly switches from a centered scheme to an upwind scheme as
the advective flux increases, so that the resulting scheme remains
stable. The drawback of this method is that it introduces (as any upwind scheme) some numerical diffusion~\cite{Fletcher1991}, which can alter the
cluster distribution by enhancing the effect of the diffusive term
$D(x,t)$~\cite{Ozkan2000,Jourdan2014}. If one is interested only in the zeroth and first moments of
the distribution (corresponding to the density and average size of
clusters respectively) or if the distribution is not strongly peaked,
the Chang-Cooper scheme can be used and yields accurate results~\cite{Jourdan2014}. In other cases, this scheme should be
used with caution.

\subsection{Coupling the master equation and the Fokker-Planck limit}

Another aspect to consider while using a discretization of the
Fokker-Planck equation~\eqref{eq-fokker-planck} is that it is valid only for $n \gg 1$. For
small values of $n$, only the master equation~\eqref{eq-recd-nsup2}--\eqref{eq-recd-neq1} can accurately describe
the evolution of clusters. This is especially true given that binding
energies $F_n^b$ (and so the parameters $\alpha_n$ through~\eqref{eq-alphan}) often strongly vary for small
cluster sizes, and that cluster distributions are known to be very
sensitive to these parameters~\cite{Johnson1978}. This observation led
to the development of hybrid
schemes~\cite{Ghoniem1980,Turkin2006,Jourdan2014}, which use the
master equation for small clusters (in the so-called ``discrete
region'') and the Fokker-Planck equation for larger clusters
(``continuous region''). A key problem is the coupling between the
discrete and continuous regions. For this purpose, it is interesting
to note that in the continuous region, a centered scheme with a mesh
size equal to~1 reduces to the original master equation when only
monomers are mobile~\cite{Turkin2006}. By imposing a centered scheme
and a mesh size equal to one at the boundary between the discrete and
continuous regions, it is thus possible to seamlessly couple the ODE~\eqref{eq-recd-nsup2}--\eqref{eq-recd-neq1}
to the PDE~\eqref{eq-fokker-planck}. If small clusters are mobile, correction terms should be
added to obtain a rigorous coupling which ensures the continuity of cluster fluxes at the interface between the two regions. It has
been shown that, with this method, the quantity of matter $S$ is conserved up to
machine precision~\cite{Jourdan2014}. This type of coupling is therefore used in all
calculations performed in the present work.

\section{Discretization of the Fokker-Planck equation}
\label{sec:num-scheme-FP}

To discretize the Fokker-Planck equation, we decouple time and space. Equations are first discretized in space, and the system of ODEs is next solved using an appropriate ODE solver. Indeed, RECD equations are in general very stiff, so we want to take advantage of the implicit solvers \cite{Hindmarsh2005,Hairer1999} which have proved to be efficient to solve such equations~\cite{Jourdan2014}. In the following, we use the CVODES package~\cite{Hindmarsh2005}, which contains variable-order and variable-step
linear multistep methods with backward differentiation formulas. As discussed in the next sections, most spatial discretizations of the advection term are built specifically to ensure total variation diminishing (TVD) properties for the forward Euler scheme,  or for any Runge-Kutta scheme which can be decomposed into successive explicit Euler stages. The TVD property ensures that an initially monotone solution remains monotone at future time steps, and does not become oscillatory. Although it cannot be guaranteed that CVODES is as precise and efficient as TVD solvers for the sole advection term, it was shown to give very satisfactory results with much lower computation times than TVD solvers for all the tests we performed.

\begin{figure}[htbp]
  \centering
  \includegraphics{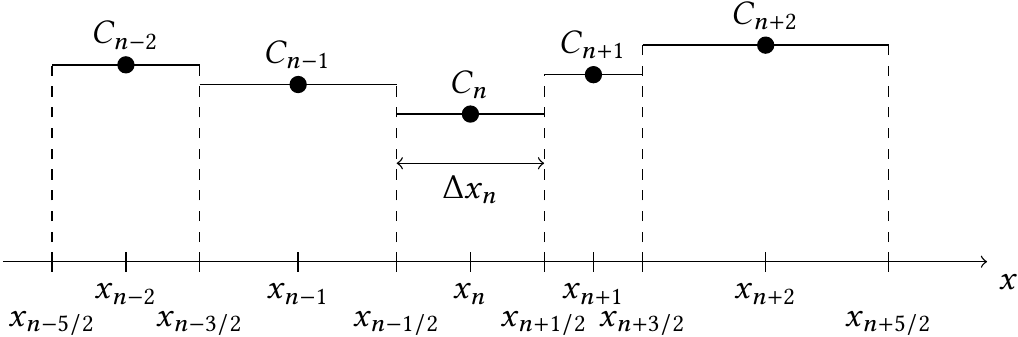}
  \caption{One-dimensional non-uniform mesh}
  \label{fig-mesh}
\end{figure}

Let us now focus on the spatial discretization of the Fokker-Planck equation. We consider the non-uniform mesh of Fig.~\ref{fig-mesh}, where $C_n$ represents the average of $C(x,t)$ over the cell of width $\Delta x_n$ (for the sake of clarity, we omit the time dependence in the following). To conserve the flux of clusters, we discretize the conservative form~\eqref{eq-FP-with-fluxes}:
\begin{equation}
  \label{eq-conservative-form}
  \frac{\mathrm{d}C_n}{\mathrm{d}t} = - \frac{J_{n+1/2}-J_{n-1/2}}{\Delta x_n},
\end{equation}
where $J_{n+1/2}$ is the flux of clusters between classes $n$ and $n+1$. It is the sum of the diffusion term $J_{n+1/2}^{\mathrm{d}}$ and the advective term $J_{n+1/2}^{\mathrm{a}}$.

The discretization of the diffusion term~\eqref{eq-Jd} can be done with the classical centered scheme:
\begin{equation}
  \label{eq-three-point-scheme-diffusion}
  J_{n+1/2}^{\mathrm{d}} = \frac{D_{n+1}C_{n+1}-D_{n}C_n}{\frac{\Delta x_n}{2}+\frac{\Delta x_{n+1}}{2}},
\end{equation}
with $D_n = D(x_n,t)$.

As mentioned previously, the discretization of the advection flux is
more difficult. Various numerical schemes offer a better accuracy than
the upwind scheme while avoiding oscillations produced by the centered
scheme. Among these schemes, which are generally called
``high-resolution schemes''~\cite{Leveque}, we only consider the
semi-discrete ones. In addition, although $J^{\mathrm{a}}$ depends linearly on
$C(x,t)$ in the present case, it can be nonlinear in general, since the drift $F$ can also
depend on cluster concentrations~\cite{Brailsford1976}. It is therefore important to use a
scheme that can also be extended to nonlinear equations.

One particularly interesting semi-discrete scheme adapted for
nonlinear equations is the central-upwind scheme developed by
Kurganov, Noelle and Petrova (KNP)~\cite{Kurganov2001}. In the wake of
the scheme proposed by Kurganov and Tadmor for nonlinear conservation
laws~\cite{Kurganov2000}, it has the advantage of being a
Riemann-solver-free approach, thus leading to simple implementation.
With the KNP scheme, the advective flux is given by
\begin{multline}
  \label{eq-KNP-flux}
  J_{n+1/2}^{\mathrm{a}} = \frac{a^+_{n+1/2} \ J^{\mathrm{a}}\left(x_{n+1/2},C^-_{n+1/2}\right)-a^-_{n+1/2} \ J^{\mathrm{a}}\left(x_{n+1/2},C^+_{n+1/2}\right)}{a^+_{n+1/2}-a^-_{n+1/2}} \\ 
+ \frac{a^+_{n+1/2} \ a^-_{n+1/2}}{a^+_{n+1/2}-a^-_{n+1/2}}\left(C^+_{n+1/2}-C^-_{n+1/2} \right).
\end{multline}
In this expression, $C^-_{n+1/2}$ and $C^+_{n+1/2}$ are the left and
right values of the reconstruction of $C$ at point $x_{n+1/2}$ from the values
$\{ C_j \}_{j \in \NN}$. Possible definitions for $C^-_{n+1/2}$ and $C^+_{n+1/2}$ will be detailed in Sections~\ref{sec:minmod-reconstruction}, \ref{sec:MC-reconstruction} and~\ref{sec:mp5} below. The reconstruction procedure is at the heart of the numerical method. 

\medskip

Before moving on to that reconstruction procedure, we explain how $a^+_{n+1/2}$ and $a^-_{n+1/2}$ are defined. In the current case, $C$ is a scalar. In the general case where $C \in \RR^N$, we denote by $\lambda_1^- < \dots <\lambda_N^-$ the $N$ eigenvalues of the Jacobian matrix $\partial J^{\mathrm{a}}/\partial C(C^-_{n+1/2})$ and $\lambda_1^+ < \dots <\lambda_N^+$ the $N$ eigenvalues of the Jacobian matrix $\partial J^{\mathrm{a}}/\partial C(C^+_{n+1/2})$. Then the coefficients $a^+_{n+1/2}$ and $a^-_{n+1/2}$ read 
\begin{align}
  \label{eq-coeffs-ap}
  a^+_{n+1/2} &= \max{\left(\lambda_N^-,\lambda_N^+,0\right)}, \\
  \label{eq-coeffs-am}
  a^-_{n+1/2} &= \min{\left(\lambda_1^-,\lambda_1^+,0\right)}.
\end{align}
It should be noted that for the (linear) problem under consideration, it holds
\begin{align}
  \label{eq-coeffs-ap-here}
  a^+_{n+1/2} &= \max{\left(F(x_{n+1/2}),0\right)}, \\
  \label{eq-coeffs-am-here}
  a^-_{n+1/2} &= \min{\left(F(x_{n+1/2}),0\right)},
\end{align}
so for a positive advective term ($F(x) > 0$), the flux simply is
\begin{equation}
  \label{eq-scheme-reduction-ap}
  J_{n+1/2}^{\mathrm{a}} = J^{\mathrm{a}}\left(x_{n+1/2},C^-_{n+1/2}\right) = F(x_{n+1/2}) C^-_{n+1/2}.
\end{equation}
Similarly, for a negative advective term, the flux reduces to
\begin{equation}
  \label{eq-scheme-reduction-ap-neg}
  J_{n+1/2}^{\mathrm{a}} = J^{\mathrm{a}}\left(x_{n+1/2},C^+_{n+1/2}\right) = F(x_{n+1/2}) C^+_{n+1/2}.
\end{equation}
In the linear case, this expression is the natural upwind flux expression. Different options are available to compute reconstructed left and right values $C^-_{n+1/2}$ and $C^+_{n+1/2}$. The goal of reconstruction is to increase the order of the method in smooth regions while retaining stability, by avoiding oscillations near discontinuities or strong gradients. 
Various reconstructions are presented in the following sections.

\subsection{Minmod reconstruction}
\label{sec:minmod-reconstruction}

The minmod reconstruction is a simple reconstruction, which
guarantees that the KNP scheme is total variation diminishing (TVD)
when the forward Euler scheme is used to discretize the left-hand side of~\eqref{eq-conservative-form}, see~\cite{Kurganov2000}. As a
consequence, an initially monotone solution remains monotone at
future time steps, and does not become oscillatory.

For a non-uniform mesh, the minmod reconstruction can be written as (see~\cite{Leveque})
\begin{align}
  \label{eq-reconstruction-slope-limiter-1}
  C^+_{n+1/2} &= C_{n+1} - \frac{\Delta x_{n+1}}{2} \sigma_{n}, \\
  \label{eq-reconstruction-slope-limiter-2}
  C^-_{n+1/2} &= C_n + \frac{\Delta x_n}{2} \sigma_{n},
\end{align}
where $\sigma_n$ is a so-called \emph{slope limiter}. It is given here by
\begin{equation}
  \label{eq-slope-limiter-minmod}
  \sigma_n = \minmod{\left(\frac{C_n-C_{n-1}}{\frac{\Delta x_n}{2} + \frac{\Delta x_{n-1}}{2}} \ , \ \frac{C_{n+1}-C_{n}}{\frac{\Delta x_{n+1}}{2} + \frac{\Delta x_n}{2}} \right)},
\end{equation}
where, for $M$ arguments, the $\minmod{}$ operator reads
\begin{equation}
  \label{eq-minmod}
  \minmod{(r_1, \dots, r_M)} = \begin{cases}
    \min \{r_1,\dots,r_M \} & \mbox{if $r_i > 0$ for all $i = 1,\dots,M$}, \\
    \max \{r_1,\dots,r_M \} & \mbox{if $r_i < 0$ for all $i = 1,\dots,M$}, \\
    0 & \textrm{otherwise.}
  \end{cases}
\end{equation}
This simple reconstruction performs in general much better than the
upwind scheme (which corresponds to $\sigma_n = 0$ everywhere), but it
reduces the slope rather strongly and therefore tends to smear out
solutions. Near extrema, as all TVD schemes, it reduces to a first order
scheme (upwind scheme) and the solution is in general smoothed.

\subsection{Monotonized central-difference reconstruction}
\label{sec:MC-reconstruction}

The monotonized central-difference (MC) reconstruction follows the
slope limiter approach. It is given by
Eqs.~\eqref{eq-reconstruction-slope-limiter-1}--\eqref{eq-reconstruction-slope-limiter-2}
where $\sigma_n$ is now given by
\begin{equation}
  \label{eq-slope-limiter-mc}
  \sigma_n = \minmod{\left(\frac{C_{n+1}-C_{n-1}}{\frac{\Delta x_{n-1}}{2}+\Delta x_n + \frac{\Delta x_{n+1}}{2}} \ , \ 2\frac{C_n-C_{n-1}}{\frac{\Delta x_n}{2} + \frac{\Delta x_{n-1}}{2}} \ , \ 2\frac{C_{n+1}-C_{n}}{\frac{\Delta x_{n+1}}{2} + \frac{\Delta x_n}{2}} \right)}.
\end{equation}

As the minmod reconstruction, it can be shown that MC reconstruction
is TVD when the explicit Euler scheme is used for the time integration~\cite{Kurganov2000}, so there can be loss of accuracy near
extrema. However, since the slopes are less reduced than with the
minmod reconstruction, it leads in general to more accurate results.

\subsection{MP5 reconstruction}
\label{sec:mp5}

To avoid smoothing of the solution at extrema, it is necessary to relax the TVD
constraints in these regions. This is the case for the MP5
reconstruction of Suresh and Huynh~\cite{Suresh1997}, which is monotonicity preserving
(MP). In particular initial upper and lower bounds on the solution are preserved at later times. 
The value $C_{n+1/2}^+$ is defined using the five values $C_{n-2}$,
$C_{n-1}$, $C_n$, $C_{n+1}$ and $C_{n+2}$ (Fig.~\ref{fig-mesh}). Essentially, the use of
five points instead of three points as in the two previous methods
makes it possible to distinguish a discontinuity (where slope
limiting should be active) from an extremum (where slope limiting leads
to excessive smearing and should not be used). Though not TVD, MP5
reconstruction does not produce oscillatory solutions when a forward Euler scheme is used to discretize the left-hand side of~\eqref{eq-conservative-form}. Any Runge-Kutta method made of successive applications of explicit Euler integrations in time may also be used.

In~\cite{Suresh1997}, this scheme was derived for a uniform
mesh. It can however be generalized to non-uniform meshes. We have built a
non-uniform variant of the scheme (see~\ref{sec:MP5-reconstruction}) and only found improvements in terms
of accuracy when using meshes exhibiting very strong variations. For the spatial meshes
considered in Sec.~\ref{sec:results-and-discussion} to discretize the
Fokker-Planck equation, no difference is visible between the results obtained by the two
variants of the scheme. It should also be noted that the non-uniform scheme is
slightly more complicated in its formulation and slightly less
efficient. Therefore, in the following, we use the scheme as derived for uniform meshes although
our meshes are not uniform.

\section{Results and discussion}
\label{sec:results-and-discussion}

We consider two examples in this section. The first one is the
evolution of quenched-in vacancies (see
Section~\ref{sec:cluster-quenched-vacancies}), which provides an
interesting setup to test the accuracy of the numerical schemes. In
particular, it allows us to validate the quality of the MP5
reconstruction.  We next use in Section~\ref{sec:loop-coarsening} the
scheme with MP5 reconstruction to test the sensitivity of cluster
distributions to the absorption and emission terms $\beta_n$ and
$\alpha_n$. More specifically, many experimental studies have focused
on the cluster distributions during coarsening by Ostwald
ripening~\cite{Ardell1988}, following the seminal theoretical work by
Lifshitz and Slyozov~\cite{Lifshitz1961} and Wagner~\cite{Wagner1961}
(LSW) for three-dimensional particles and by
Kirchner~\cite{Kirchner1973} and Burton and Speight~\cite{Burton1986}
(KBS) for two-dimensional dislocation loops. We study here the
influence of some parameters on the cluster distribution shapes during
the coarsening of dislocation loops, a simulation which is impossible
by a direct integration of the original ODE set of RECD.

\subsection{Clustering of quenched-in vacancies}
\label{sec:cluster-quenched-vacancies}

We consider Eqs.~\eqref{eq-recd-nsup2}--\eqref{eq-recd-neq1} with the following parameters:
\begin{align}
  \label{eq-koiwa}
  \beta_n &= \beta \qquad \mbox{for } n \ge 2, \\
  \beta_1 &= \eta \beta, \\
  \alpha_n &= 0 \qquad \mbox{for } n \ge 2,
\end{align}
for some $\eta < 1$. The initial condition corresponds to a material with quenched-in vacancies: $C_1(t=0) = C_{\mathrm{q}}$ and $C_n(t=0) = 0$ for any $n \ge 2$. 

This test case was first proposed by Koiwa~\cite{Koiwa1974} to
investigate the validity of the grouping method of Kiritani~\cite{Kiritani1973}, and was
subsequently used to validate the grouping method of Golubov \emph{et
  al.} (G-method)~\cite{Golubov2001}. Since clusters cannot dissociate
($\alpha_n = 0$), the ratio of the advective flux to the diffusive
flux is high (see Eqs.~\eqref{eq-Fx} and~\eqref{eq-Dx}), which makes
this test particularly interesting to validate the numerical schemes
for the Fokker-Planck equation, and in particular to investigate
numerical diffusion. As time increases, the concentration of monomers
$C_1$ decreases due to the absorption reactions, until it becomes zero. At this time, the cluster
distribution does not evolve any more. 

To perform this test, we simply choose $\beta = a_0 D_1$, where $a_0$ is the lattice parameter and $D_1$ is the diffusion coefficient (Eq.~\eqref{eq-diffusion-coefficient-monomer}). Numerical values of parameters are given in Tab.~\ref{tab-parameters-koiwa}. The value of $\beta$ only has an influence on the time at which the distribution reaches its final stationary state~\cite{Koiwa1974}. This time is set to $2\times 10^{5}$~s in our calculations.

\begin{table*}[htb]
  \centering
  \begin{tabular}{lllll}
    \hline
    Symbol & Description & Value & Unit \\ \hline
    $a_0$ & Lattice parameter & 0.364 & nm \\
    $V_{\mathrm{at}} = a_0^3/4$ & Atomic volume & $1.206\times 10^{-29}$ & m$^3$ & \\
    $E^{\mathrm{m}}_1$ & Migration energy of vacancies & 0.7 & eV \\
    $D_{0,1}$       & Diffusion prefactor of vacancies & $2\times 10^{-6}$ & m$^2$ s$^{-1}$ \\
    $T$           & Temperature        & 523 & K \\
    $C_{\mathrm{q}} $ & Initial concentration of vacancies & $10^{-8}/V_{\mathrm{at}}$ & m$^{-3}$ \\ 
    $\eta$ & Coefficient for the clustering of two vacancies & $10^{-4}$ & \\  \hline
  \end{tabular}
\caption{Parameters used for the simulation of clustering of quenched-in vacancies~\cite{Golubov2001}.}
\label{tab-parameters-koiwa}
\end{table*}

\begin{figure*}[htb]
  \centering
  \includegraphics[width=\textwidth]{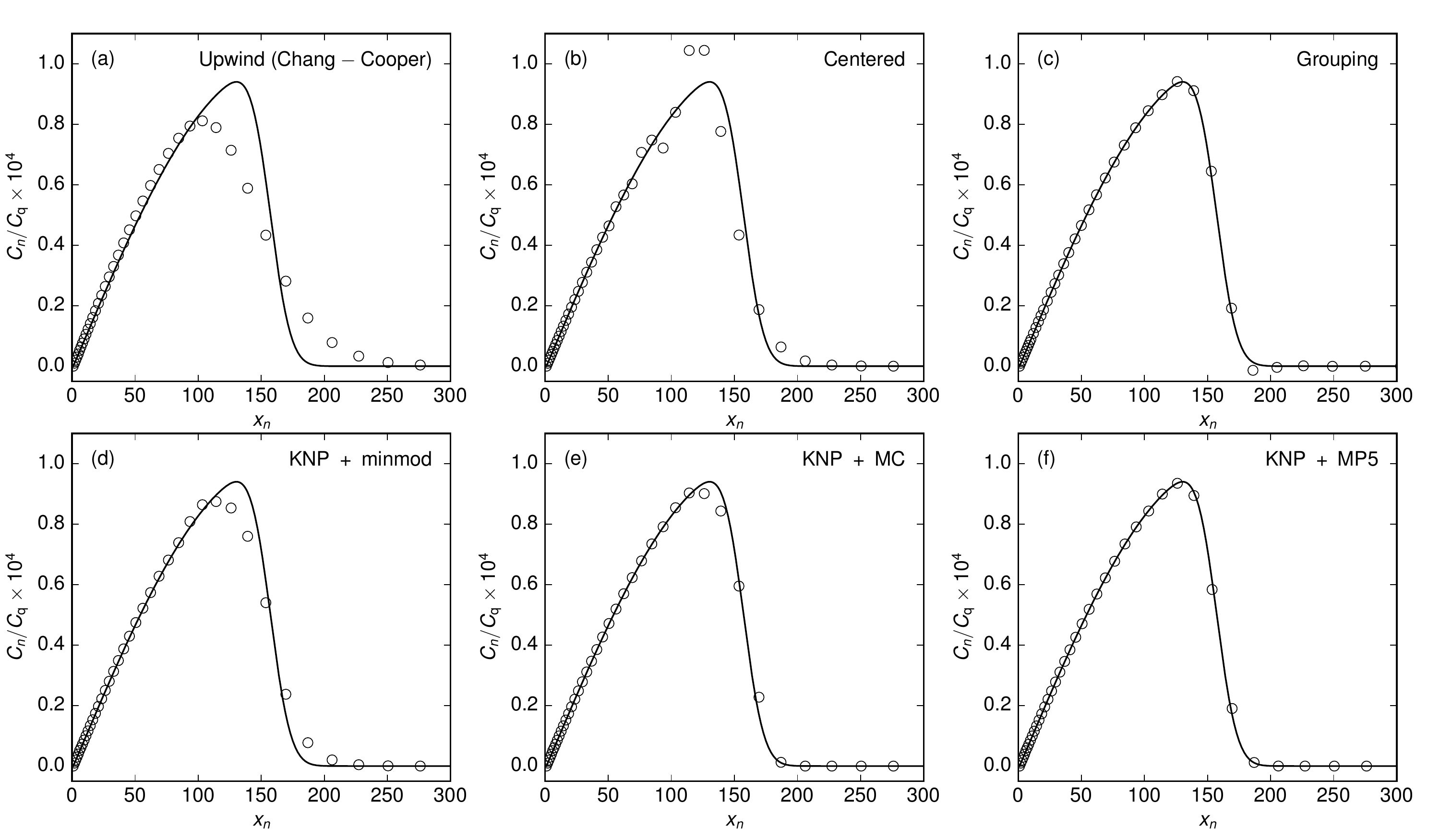}
  \caption{Long time limit cluster size distributions for the test proposed
    by Koiwa~\cite{Koiwa1974} (clustering of quenched-in vacancies)
    using different approximations (circles): (a) Chang-Cooper scheme (upwind
    scheme), (b) centered scheme, (c) grouping method (G-method), KNP
    with (d) minmod, (e) MC and (f) MP5 reconstructions. The exact
    calculation, taking into account all cluster classes, is shown by
    the black solid line.}
  \label{fig-koiwa}
\end{figure*}

In Fig.~\ref{fig-koiwa} we show the stationary cluster distributions 
obtained for $\eta = 10^{-4}$ with different numerical schemes: the Chang-Cooper scheme~\cite{Chang1970},
which reduces to the upwind scheme in this case ($F_n = F(x_n)$):
\begin{equation}
  \label{eq-upwind}
  J^{\mathrm{a}}_{n+1/2} = F_nC_n,
\end{equation}
the centered scheme:
\begin{equation}
  \label{eq-centered-scheme}
  J^{\mathrm{a}}_{n+1/2} = \frac{1}{2}\left(F_nC_n + F_{n+1}C_{n+1} \right),
\end{equation}
and the KNP scheme with the three reconstructions presented in Sections~\ref{sec:minmod-reconstruction}, \ref{sec:MC-reconstruction} and~\ref{sec:mp5}. For the sake of completeness, we also present the
results obtained by the G-method, which is not based on the
Fokker-Planck equation but on grouping of equations of the initial
master equation. The reference calculation, shown by the black solid
line in each figure, is produced by including all cluster classes in
the simulation. The mesh used for the other calculations is constant over
the 4 first cluster classes and then increases geometrically,
that is $\Delta x_{n+1} = (1+\epsilon) \Delta x_n$ with $\epsilon =
0.1$ for $n \geq 4$. Near the distribution peak, the width of the mesh is close to 7.

As expected, the upwind method (Fig.~\ref{fig-koiwa}-a) introduces
substantial numerical diffusion and the distribution broadens. It
should be noted, however, that the zeroth and first moments of the
distribution are well reproduced. The error on the cluster density ($n
\ge 2$) is about 5~\%, while for the average number of vacancies
inside clusters it is only 2~\%. Owing to its simplicity and high
efficiency (see below), this method is attractive if the quantity of interest is
the cluster density or the average number of vacancies. However, if a
precise description of the cluster distribution is needed, this scheme
should be avoided. The centered scheme (Fig.~\ref{fig-koiwa}-b)
introduces spurious oscillations in the solution. It is therefore
clearly inappropriate for this test, and more generally for all cases
with strong advection terms. Significant improvement is obtained with
the simple minmod reconstruction (Fig.~\ref{fig-koiwa}-d): no
oscillations appear and it produces less numerical diffusion than the
upwind scheme. Numerical diffusion can be even more reduced by the use
of MC reconstruction (Fig.~\ref{fig-koiwa}-e), although near the peak
one can still observe a small deviation with respect to the reference
calculation. Among the reconstructions that were tested, MP5 clearly
appears to be the best one; it perfectly reproduces the reference
calculation (Fig.~\ref{fig-koiwa}-f). Incidentally, this calculation
also shows that the error due to the transformation of the master
equation into a Fokker-Planck equation is completely
negligible. Although the Fokker-Planck approximation is derived from a Taylor expansion of the master equation, which is strictly valid only for $n \gg 1$, we observe numerically that using this approximation for $n$ as low as 5 does not introduce error in the solution. Concerning grouping methods, the G-method
(Fig.~\ref{fig-koiwa}-c) performs very well also, the only difference
with the MP5 reconstruction being the presence of unphysical negative
concentrations in the large cluster size tail of the distribution
(value for $x_n$ around~180 in Fig.~\ref{fig-koiwa}-c). Such
unphysical effects cannot appear with the MP5 reconstruction since the
scheme is monotonicity preserving. We conclude from this test that the
MP5 scheme is indeed very accurate.

The primary purpose of this work is to compare the accuracy of the
discretization schemes. It is however instructive to compare the
computation times required to perform the simulations with a given
time solver. The wall-clock computation times on a single core
processor  are given in Tab.~\ref{tab-comput_times} using
CVODES package. To make the comparison sensible between the different
schemes, the Jacobian matrix is evaluated by CVODES using finite
differences and the standard dense linear solver included in the
package is used. In all cases, the computation time is rather short to reach $t = 2\times 10^5$~s (less than
10~s), but it varies from one method to another. The shortest times
are for the Chang-Cooper and centered schemes. It gradually increases
with the complexity of limiters. For larger systems, linear algebra
dominates the overall computation time and one can expect the
efficiency of the grouping method to decrease, since it uses two
equations per group, so approximately twice more equations than the
Fokker-Planck based methods. A systematic study of the efficiency is
however difficult to perform, since a time solver can be more or less
adapted to a spatial discretization. For example, implicit methods as
implemented in CVODES are probably not optimal when limiters are used,
due to the frequent switches in the discretization
stencils. Implicit-explicit methods, treating advective fluxes in the
explicit part, may be more efficient. We emphasize however that even
with CVODES, the computation time is reasonable with the MP5 scheme. If a
large number of calculations are required, as for parameter fitting,
it may be advantageous to use Chang-Cooper scheme or minmod limiter to
get rough solutions, and then refine the final solution with the MP5
scheme.

\begin{table}[htb]
  \centering
  \begin{tabular}{lc}
    Scheme & Computation time (s) \\ \hline
    Chang-Cooper & 1.3  \\
    Centered & 1.3 \\
    Grouping & 3.0 \\
    KNP + minmod & 2.8 \\
    KNP + MC & 4.2 \\
    KNP + MP5 & 9.1 \\ \hline    
  \end{tabular}
  \caption{Wall-clock computation times to reach $t=2\times 10^5$~s with various numerical schemes, for the test proposed by Koiwa~\cite{Koiwa1974}. Time integration is performed on a 2.80 GHz processor using CVODES, with finite difference calculation of the Jacobian matrix and the dense linear solver included in the package.}
  \label{tab-comput_times}
\end{table}

\subsection{Loop coarsening}
\label{sec:loop-coarsening}

Loop coarsening by Ostwald ripening was studied theoretically by
Burton and Speight~\cite{Burton1986} and Kirchner~\cite{Kirchner1973}
(a model known as KBS). They found that the loop distribution in the coarsening regime
obeys the following law:
\begin{equation}
  \label{eq-loop-distrib-KBS}
  C(\rho,t) = A \frac{\rho}{(2-\rho)^4} \exp{\left(-\frac{4}{2-\rho}\right)} \frac{1}{\langle r(t) \rangle^2},
\end{equation}
where $\rho = r/\langle r(t) \rangle $, $r$ is the loop radius and
$\langle r(t) \rangle$ is the mean loop radius of the cluster
distribution. The parameter $A$ is a normalizing constant, which can
be expressed, for example, as a function of the total concentration of
defects inside the loops, which remains constant during the coarsening
process. For a pure prismatic loop, the loop radius $r$ is related to
the number of particles $x$ by $\pi r^2 b = x V_{\mathrm{at}}$, where $b$ is the Burgers vector.

To make the problem tractable, in addition to the classical hypotheses
of the LSW theory for 3D particles~\cite{Lifshitz1961,Wagner1961}, two main approximations are made
in the KBS approach in order to obtain Eq.~\eqref{eq-loop-distrib-KBS}:
\begin{enumerate}
\item The formation energy of unfaulted, circular loops is
  given by a simple line tension model:
    \begin{equation}
      \label{eq-line-tension-energy}
      E^{\mathrm{f}}(r) = 2\pi r \mathcal{T},
    \end{equation}
  where $\mathcal{T}$ is the line tension, which is considered
  constant. Another expression is given by Hirth and
  Lothe~\cite{Hirth1968} using isotropic elasticity, for an isolated
  pure prismatic loop:
  \begin{equation}
    \label{eq-energy-pure-prismatic}
    E^{\mathrm{f}}(r) = 2\pi r \frac{\mu b^2}{4\pi(1-\nu)} \left[\ln{\left(\frac{4r}{r_{\mathrm{c}}}\right)}-1\right],
  \end{equation}
  where $\mu$ is the shear modulus, $\nu$ is Poisson's ratio and
  $r_{\mathrm{c}}$ is a core cut-off radius. Since $F_n^{\mathrm{b}} = E^{\mathrm{f}}_1 +
E^{\mathrm{f}}_{n-1} - E^{\mathrm{f}}_{n}$ with $E^{\mathrm{f}}_n = E^{\mathrm{f}}(r_n)$ in Eq.~\eqref{eq-alphan}, the logarithmic dependency may affect the
  emission rates $\alpha_n$ and therefore
  the shape of the loop distribution. 

\item The diffusion flux of defects to a loop can be written as the
  flux to a spherical cluster of the same radius, which means that
  $\beta_n$ is given by Eq.~\eqref{eq-betan} with $\pi r_n^2 b = x_n
  V_{\mathrm{at}}$ (the loop is assumed to be pure
  prismatic). Therefore, the toroidal shape of the cluster is not
  taken into account. Seeger and G\"osele~\cite{Seeger1977} have shown that, when the
  diffusion problem is solved in toroidal geometry, the absorption
  coefficient reads
  \begin{equation}
    \label{eq-betan-seeger-gosele}
    \beta_n = 2\pi r_n \frac{2\pi}{\ln{\left(\frac{8r_n}{r_{\mathrm{p}}}\right)}} D_1,
  \end{equation}
where $r_{\mathrm{p}}$ is a core cut-off radius. More accurate 
  formulas take into account the drift-diffusion process due to the
  elastic interaction between the defect and the loop~\cite{Dubinko2005,Jourdan2015}.
\end{enumerate}

To test the validity of the formula~\eqref{eq-loop-distrib-KBS} given by the KBS theory and the
sensitivity of cluster distributions to absorption rates $\beta_n$ and
to the energetics of loops, RECD calculations were performed with
different approximations:
\begin{enumerate}
\item[(a)] Spherical model for the absorption kinetics
  (Eq.~\eqref{eq-betan}), line tension model for the energetics of
  loops (Eq.~\eqref{eq-line-tension-energy}), as for the KBS approach.
\item[(b)] Spherical model for the absorption kinetics
  (Eq.~\eqref{eq-betan}), formula based on isotropic elasticity for
  the energetics of loops (Eq.~\eqref{eq-energy-pure-prismatic}).
\item[(c)] Toroidal model for the absorption kinetics
  (Eq.~\eqref{eq-betan-seeger-gosele}), line tension model for the
  energetics of loops (Eq.~\eqref{eq-line-tension-energy}).
\end{enumerate}

The list of parameters corresponding to loops in aluminum is given in
Tab.~\ref{tab-parameters-aluminum}. In our calculations the line
tension is given by $\mathcal{T} = \kappa \mu b^2$, where $\kappa$ has
been adjusted to give a continuous evolution for binding energies from
di-vacancies, the value of which is imposed, to larger clusters, for
which the line tension is used. Likewise, the value of the core
cut-off radius $r_{\mathrm{c}}$ in Tab.~\ref{tab-parameters-aluminum}
ensures a smooth transition between di-vacancies and larger loops.

\begin{table*}[htb]
  \centering
  \begin{tabular}{lllll}
    \hline
    Symbol & Description & Value & Unit & Reference \\ \hline
    $V_{\mathrm{at}}$ & Atomic volume & $1.648\times 10^{-29}$ & m$^3$ & \\
    $b$           & Burgers vector & 0.2857 & nm \\
    $E^{\mathrm{m}}_1$ & Migration energy of vacancies & 0.61 & eV & \cite{Ehrhart1991}\\
    $D_{0,1}$       & Diffusion prefactor of vacancies & $1.18\times 10^{-5}$ & m$^2$ s$^{-1}$ & \cite{Bako2011} \\
    $T$           & Temperature        & 600 & K \\
    $\mu$        & Shear modulus       & 26.5 & GPa & \cite{Bako2011} \\
    $\nu$ & Poisson's ratio & 0.345 & & \cite{Bako2011} \\
    $E^{\mathrm{f}}_1$ & Formation energy of vacancies & 0.67 & eV & \cite{Ehrhart1991}\\
    $F^{\mathrm{b}}_2$ & Binding energy of two vacancies & 0.2 & eV & \cite{Ehrhart1991}\\ 
    $\kappa$      & Coefficient for the line tension  & 0.1 & \\
    $r_{\mathrm{c}}$ & Core cut-off radius for the elastic law & 0.15 & nm & \cite{Bako2011}\\
    $r_{\mathrm{p}}$ & Core cut-off radius for the absorption rate & 0.5713 & nm & \\ \hline
  \end{tabular}
\caption{Parameters used for the simulation of vacancy loop coarsening in aluminum.}
\label{tab-parameters-aluminum}
\end{table*}

\begin{figure*}[htbp]
  \centering
  \includegraphics[width=\textwidth]{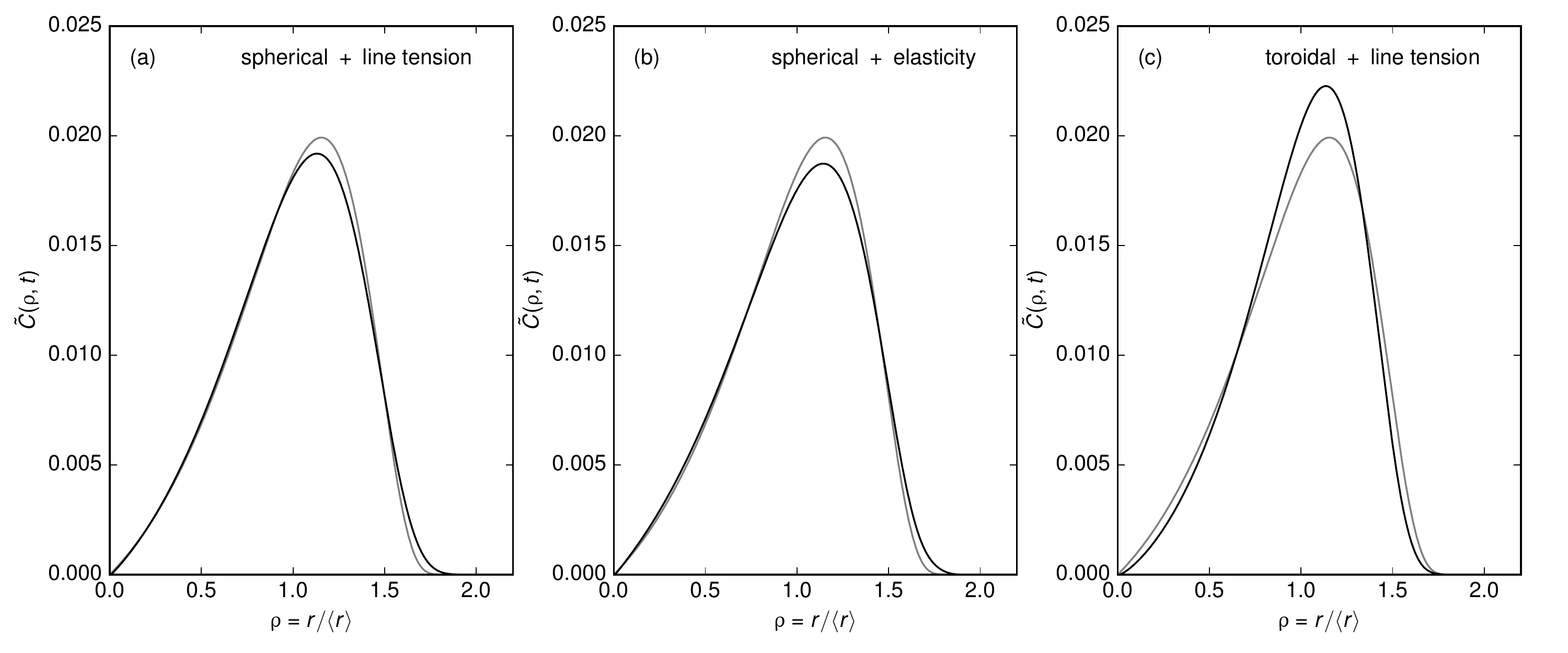}
  \caption{Distributions obtained by RECD using various approximations
    (black line, see text) and given by the KBS theory (gray line).}
  \label{fig-distrib-kbs}
\end{figure*}

Initial distributions are only populated for the class $n = 100$ ($x_n
\approx 391$ due to the geometric progression of the mesh), with a
concentration equal to $10^{23}$~m$^{-3}$. This initial condition
permits to determine the normalizing constant $A$
in~\eqref{eq-loop-distrib-KBS}. The use of a Fokker-Planck approach
enables us to simulate cluster distributions with less than 1000
classes, whereas a simulation with all cluster classes would require
approximately 16 million equations. The first 20 cluster classes are
described by the master equation, then a geometric progression is used
with $\epsilon = 0.03$ for 100 classes and $\epsilon$ is decreased to
0.01 for the remaining classes.  Calculations are performed with MP5
reconstruction up to a time at which the normalized distributions are
stationary. It has been checked that it is the case for all
parametrizations at $t=5$~s. In Fig.~\ref{fig-distrib-kbs} (a)--(c) we
show the normalized cluster distributions $\tilde{C}(\rho,t) =
C(\rho,t) \langle r(t) \rangle^2/A$ obtained by RECD within the
Fokker-Planck approach at $t = 5$~s as well as the one predicted by
the KBS theory.

With parametrization (a), which is based on the same approximations as
KBS theory, a good agreement is obtained
(Fig.~\ref{fig-distrib-kbs}-(a)). The profile given by RECD is
slightly less peaked and broader than the theoretical profile. We have
checked, by varying the mesh size, that numerical diffusion is
negligible and cannot explain this small discrepancy. Most probably it
comes from the probabilistic nature of the growth law in RECD, whereas
in KBS it is purely deterministic. In RECD, growth occurs by successive absorptions of monomers but emissions can also occur, whereas in KBS theory any cluster of size larger than the critical radius will grow. Such an argument has been put
forward to explain the discrepancy between LSW theory and RECD
calculations of the coarsening of 2D
deposits~\cite{Berthier2011}. Broader distributions were also obtained
with RECD, but the difference was much more important between LSW and
RECD than in the present case. Our results are also in line with the
work by Senkov~\cite{Senkov2008}, who pointed out the role of
fluctuations to explain why experimental distributions are often
broader than those predicted by LSW theory.

When Eq.~\eqref{eq-energy-pure-prismatic} based on isotropic
elasticity is used, results are almost identical to those obtained
with the simple line tension model
(Fig.~\ref{fig-distrib-kbs}-(b)). This result is compatible with
dislocation dynamics simulations of the coarsening of
loops~\cite{Bako2011}: the authors obtained distributions in good
agreement with KBS theory, although no assumption on the energetics of
loops as in Eqs.~\eqref{eq-line-tension-energy} and
\eqref{eq-energy-pure-prismatic} is made in this type of
simulations. These results can also be compared to experimental
results in silicon.  Some authors found that the loop distributions
followed a KBS law~\cite{Bonafos1998}, while others found loop
distributions with much longer tails~\cite{Pan1997}. They ascribed
this discrepancy to the logarithmic term in
Eq.~\eqref{eq-energy-pure-prismatic}. Although our simulations are
performed on a different material, they clearly do not support this
interpretation.

Interestingly, a more significant difference appears between RECD
results and KBS theory if the absorption coefficient takes into
account the toroidal geometry of the loop
(Fig.~\ref{fig-distrib-kbs}-(c)). In this case, the distribution is
sharper. This can be explained by the logarithmic term in
Eq.~\eqref{eq-betan-seeger-gosele} which decreases the absorption rate
for large clusters with respect to the spherical approximation,
thereby lowering their growth rate. 

It should be noted that the influence of other
loops on the absorption flux due to overlapping diffusion fields and
on the energetics of loops is not taken into account in these calculations. More complex
models would be necessary to include these effects, see \emph{e.g.}
\cite{brailsford_dependence_1979}, but this is beyond the scope of the
present work. What we want to emphasize here is the importance of
using a proper numerical scheme to obtain physical distributions which
are not distorted by numerical diffusion, in order to compare the
various physical models. Essentially, our conclusions would have been
difficult, if not impossible to draw, with other numerical schemes.

\section{Conclusion}
\label{sec:conclusion}

In this work we have introduced a numerical scheme to accurately and
efficiently solve cluster dynamics equations. It is based on a coupling
between the master equation (for small clusters) and the
discretization of the Fokker-Planck equation (for large clusters). To
avoid the distortion of cluster distributions in the region where the
Fokker-Planck equation is used, special care must be taken. We have
shown that high resolution methods must be used for the advection
term. More specifically, the KNP scheme coupled
with the MP5 reconstruction appears to be a very attractive method. On the
cases that have been tested, the
results obtained by MP5 reconstruction are in very good agreement with
the reference solution even when using rather large mesh sizes. The so-obtained numerical method compares favourably with the grouping
methods, since it uses twice less equations and prevents the appearance of
negative concentrations. In addition, the use of the Fokker-Planck
equation permits to clearly identify the drift and diffusion terms,
which can be interesting for physical interpretation. The method can be easily generalized to simulations containing two species by applying the algorithm along each dimension, except if cross-terms in the Fokker-Planck equation  are present~\cite{Jourdan2014}. Such terms arise if clusters containing several species are mobile. For higher dimensions, stochastic methods are probably more appropriate due to the increase in the number of equations to solve~\cite{Marian2011,Gherardi2012,Dunn2015}. It should also be noted that the method is not adapted if large clusters are mobile since it is based on a Taylor expansion, whereas grouping
methods are applicable to this case~\cite{Golubov2007}. 

Using this method, we have simulated the loop coarsening in aluminum
and compared our results to the KBS theory. When
the same assumptions are taken for the absorption coefficients and
energetics of loops, a good agreement is obtained. Cluster
distributions are found to be only slightly broader in RECD
calculations, which can be ascribed to the probabilistic nature of the
growth law in RECD. Additional calculations show that normalized
cluster distributions do not depend much on the energetics of loops,
but that they are sensitive to the model used for the absorption
coefficients.

\section{Acknowledgments}
\label{sec:acknowledgments}

This work was performed within the NEEDS project MathDef ({\em Projet Blanc}). We gratefully acknowledge the financial support of NEEDS. The CRESCENDO code, which was used for this study, is developed in collaboration with EDF R\&D within the framework of I3P institute.

\appendix

\section{MP5 reconstruction}
\label{sec:MP5-reconstruction}

We recall here briefly the MP5 reconstruction by Suresh and
Huynh~\cite{Suresh1997}, adapted to the case of non-uniform meshes. We
assume that $F(x)$ is positive at $x_{n+1/2}$, so
in view of~\eqref{eq-scheme-reduction-ap} we need to determine
$C_{n+1/2}^-$. The arguments can be easily adapted to the case $F(x) < 0$.

The method proceeds in two steps that are successively detailed in~\ref{sec:original-interface-value} and \ref{sec:limiting} below:
\begin{enumerate}
\item A high order approximation of $C_{n+1/2}^-$ is calculated, based on the values $\{ C_j \}_{j \in \NN}$;
\item The value of $C_{n+1/2}^-$ is possibly changed (limiting process) to ensure that if the $\{ C_j \}_{j \in \NN}$ are initially monotone, they remain monotone at the subsequent timestep. This amounts to imposing the value of $C_{n+1/2}^-$ to lie inside some prescribed intervals which depend on the initial values of $\{ C_j \}_{j \in \NN}$. Limiting leads to a lower order reconstruction, but guarantees that no oscillations appear. Intervals can be set to impose TVD property, but it can lead to too much smearing near extrema. With MP5 reconstruction the intervals are enlarged near extrema, to keep as much as possible the initial high order approximation of $C_{n+1/2}^-$: although TVD property is lost in general, monotonicity is preserved and the accuracy is higher than with TVD schemes.
\end{enumerate}

\subsection{Original interface value}
\label{sec:original-interface-value}

The value at the interface $C_{n+1/2}^-$ is defined by
$C_{n+1/2}^- = \tilde{C}_n(x_{n+1/2})$, where
\begin{equation}
  \label{eq-def-tildeCj}
  \tilde{C}_n(x) = \sum_{i=1}^5 a_{i-1} \left(x-x_n\right)^{i-1}.
\end{equation}
Following Ref.~\cite{Capdeville2008}, the coefficients $a_i$ are obtained by imposing that
\begin{equation}
  \label{eq-conservation-assumption}
  \frac{1}{\Delta x_{n+k}}\int_{I_{n+k}} \tilde{C}_n(x) \,\mathrm{d}x = C_{n+k} \qquad \textrm{for any } k\in [-2,1,0,1,2],
\end{equation}
where $I_{n}$ is the interval corresponding to index $n$. For a uniform mesh, this leads to
\begin{equation}
  \label{eq-C-interface-uniform}
  C_{n+1/2}^{-} = \frac{1}{60}\left(2C_{n-2} - 13C_{n-1} + 47 C_{n} + 27 C_{n+1} - 3 C_{n+2}\right).
\end{equation}

Higher order reconstructions may be used~\cite{Suresh1997}, but in practice for cluster dynamics problems this reconstruction appears accurate enough. To determine $C_{n+1/2}^-$ in the non-uniform case, a linear system must be solved at each interface point for the non-uniform scheme, which makes it more complicated to implement and less efficient than the uniform scheme. For meshes with low spatial variation as those considered in the present work, we have found that the two versions lead to nearly identical results.

\subsection{Slope limiter}
\label{sec:limiting}

The following steps are performed to preserve the monotonicity of the solution:
\begin{itemize}
\item The following values are computed:
  \begin{align}
    \label{eq-c-ul}
    C_n^{\mathrm{UL}} &= C_n + 2\gamma\frac{C_n-C_{n-1}}{\Delta x_n + \Delta x_{n-1}} \Delta x_n  \\
    \label{eq-c-mp}
    C_n^{\mathrm{MP}} &=  C_n + \minmod{\left(C_{n+1}-C_n,C_n^{\mathrm{UL}}-C_n\right)},
  \end{align}
  with $\gamma \ge 2$. We adopted $\gamma=4$, in agreement with Ref.~\cite{Suresh1997}.
\item If $(C_{n+1/2}^- - C_n) (C_{n+1/2}^- - C_n^{\mathrm{MP}}) < 0$, the value of $C_{n+1/2}^{-}$ is left unchanged. Otherwise, we must perform the limiting process which is detailed in the next step.
\item If limiting is necessary, the value $C_{n+1/2}^-$ is changed to
\[\minmod{\left(C_n^{\mathrm{min}}-C_{n+1/2}^-,C_n^{\mathrm{max}}- C_{n+1/2}^-\right)},
\]
with:
  \begin{align}
    \label{eq-vmin}
    C_n^{\mathrm{min}} &= \max{\left[\min{\left(C_n,C_{n+1},C^{\mathrm{MD}}\right)},\min{\left(C_n,C_n^{\mathrm{UL}},C_n^{\mathrm{LC}}\right)}\right]} \\
    \label{eq-vmax}
    C_n^{\mathrm{max}} &= \min{\left[\max{\left(C_n,C_{n+1},C^{\mathrm{MD}}\right)},\max{\left(C_n,C_n^{\mathrm{UL}},C_n^{\mathrm{LC}}\right)}\right]} \\
    \label{eq-CjLC-explicit}
    C_n^{\mathrm{LC}} &= C_n + 2 \frac{C_n-C_{n-1}}{\Delta x_n + \Delta x_{n-1}} \frac{\Delta x_n}{2} + \frac{1}{6} \zeta_n \minmod{\left(\delta_{n-1},\delta_n\right)} \Delta x_n \left(\Delta x_n + \Delta x_{n-1} \right). \\
    \label{eq-ConcMD}
    C_n^{\mathrm{MD}} &= C_n^{\mathrm{AV}} + \minmod{\left(C_n^{\mathrm{FL}}-C_n^{\mathrm{AV}},C_n^{\mathrm{FR}}-C_n^{\mathrm{AV}}\right)}. \\
    \label{eq-ConcFL}
    C_n^{\mathrm{FL}} &= C_n + \frac{C_n-C_{n-1}}{\Delta x_n + \Delta x_{n-1}} \Delta x_n \\
    \label{eq-ConcFR}
    C_n^{\mathrm{FR}} &= C_{n+1} + \frac{C_{n+1}-C_{n+2}}{\Delta x_{n+1} + \Delta x_{n+2}} \Delta x_{n+1} \\
    \label{eq-ConcAV}
    C_n^{\mathrm{AV}} &= \frac{C_n \Delta x_{n+1} + C_{n+1} \Delta x_n}{\Delta x_{n+1} + \Delta x_n} \\  
    \label{eq-inequality-beta-general}
    \zeta_n &\le \zeta_n^{\max} \\
    \label{eq-betamax}
    \zeta_n^{\max} &= (2\gamma-1)\frac{3}{4}\frac{\Delta x_n + \Delta x_{n-2} + 2\Delta x_{n-1}}{\Delta x_n + \Delta x_{n-1}} \\
    \label{eq-delta-nmhalf}
    \delta_{n-1/2}^{\mathrm{MM}} &= \minmod{\left(\delta_{n-1},\delta_n\right)} \\
    \label{eq-second-order-derivative}
    \delta_n &= \frac{8 d_n}{(\Delta x_{n+1}+\Delta x_{n-1}+2\Delta x_n)(\Delta x_n + \Delta x_{n+1})(\Delta x_n + \Delta x_{n-1})} \\
    \label{eq-dj}
    d_n &= C_{n-1} (\Delta x_{n+1}+\Delta x_{n}) + C_{n+1}(\Delta x_n + \Delta x_{n-1}) - C_n (\Delta x_{n+1}+\Delta x_{n-1}+2\Delta x_n).
  \end{align}
  
\end{itemize}

For a uniform mesh and $\gamma = 2$, inequality~\eqref{eq-inequality-beta-general} leads to $\zeta_n \le 9/2$, which is in line with the choice $\zeta_n = 4$ in Ref.~\cite{Suresh1997} (we recall that we have set $\gamma = 4$ in this work). Reducing $\zeta_n$ activates limiting more often, thereby increasing stability at the cost of accuracy.

Finally, the following timestep restriction is necessary to ensure that the scheme is monotonicity preserving with an explicit Euler integration:
\begin{equation}
  \label{eq-timestep-restriction}
  \Delta t \le \min_n \left(\frac{1}{F_n} \frac{\Delta x_n}{1+\frac{2\gamma\Delta x_n}{\Delta x_n + \Delta x_{n-1}}} \right)
\end{equation}


\begin{thebibliography}{58}
\expandafter\ifx\csname natexlab\endcsname\relax\def\natexlab#1{#1}\fi
\providecommand{\bibinfo}[2]{#2}
\ifx\xfnm\relax \def\xfnm[#1]{\unskip,\space#1}\fi
\bibitem[{Langer and Schwartz(1980)}]{Langer1980}
\bibinfo{author}{J.~S. Langer}, \bibinfo{author}{A.~J. Schwartz},
  \bibinfo{journal}{Phys. Rev. A} \bibinfo{volume}{21} (\bibinfo{year}{1980})
  \bibinfo{pages}{948}.
\bibitem[{Maugis and Goun\'e(2005)}]{Maugis2005}
\bibinfo{author}{P.~Maugis}, \bibinfo{author}{M.~Goun\'e},
  \bibinfo{journal}{Acta Mater.} \bibinfo{volume}{53} (\bibinfo{year}{2005})
  \bibinfo{pages}{3359}.
\bibitem[{Soisson and Fu(2007)}]{soisson_cu-precipitation_2007}
\bibinfo{author}{F.~Soisson}, \bibinfo{author}{C.-C. Fu},
  \bibinfo{journal}{Phys. Rev. B} \bibinfo{volume}{76} (\bibinfo{year}{2007})
  \bibinfo{pages}{214102}.
\bibitem[{Domain et~al.(2004)Domain, Becquart, and Malerba}]{Domain2004c}
\bibinfo{author}{C.~Domain}, \bibinfo{author}{C.~S. Becquart},
  \bibinfo{author}{L.~Malerba}, \bibinfo{journal}{J. Nucl. Mater.}
  \bibinfo{volume}{335} (\bibinfo{year}{2004}) \bibinfo{pages}{121}.
\bibitem[{Jourdan et~al.(2010)Jourdan, Bocquet, and Soisson}]{Jourdan2010a}
\bibinfo{author}{T.~Jourdan}, \bibinfo{author}{J.-L. Bocquet},
  \bibinfo{author}{F.~Soisson}, \bibinfo{journal}{Acta Mater.}
  \bibinfo{volume}{58} (\bibinfo{year}{2010}) \bibinfo{pages}{3295}.
\bibitem[{Ortiz and Caturla(2007)}]{ortiz2007_2}
\bibinfo{author}{C.~J. Ortiz}, \bibinfo{author}{M.~J. Caturla},
  \bibinfo{journal}{Phys Rev B} \bibinfo{volume}{75} (\bibinfo{year}{2007})
  \bibinfo{pages}{184101}.
\bibitem[{Clouet(2009)}]{clouet_modeling_2009}
\bibinfo{author}{E.~Clouet}, in: \bibinfo{editor}{D.~U. Furrer},
  \bibinfo{editor}{S.~L. Semiatin} (Eds.), \bibinfo{booktitle}{Fundamentals of
  Modeling for Metals Processing}, volume \bibinfo{volume}{22A} of
  \textit{\bibinfo{series}{{ASM} Handbook}}, \bibinfo{publisher}{ASM
  International}, \bibinfo{year}{2009}, p. \bibinfo{pages}{203}.
\bibitem[{Jourdan et~al.(2010)Jourdan, Soisson, Clouet, and
  Barbu}]{Jourdan2010}
\bibinfo{author}{T.~Jourdan}, \bibinfo{author}{F.~Soisson},
  \bibinfo{author}{E.~Clouet}, \bibinfo{author}{A.~Barbu},
  \bibinfo{journal}{Acta Mater.} \bibinfo{volume}{58} (\bibinfo{year}{2010})
  \bibinfo{pages}{3400}.
\bibitem[{Jourdan and Crocombette(2012)}]{Jourdan2012}
\bibinfo{author}{T.~Jourdan}, \bibinfo{author}{J.-P. Crocombette},
  \bibinfo{journal}{Phys. Rev. B} \bibinfo{volume}{86} (\bibinfo{year}{2012})
  \bibinfo{pages}{054113}.
\bibitem[{Hairer and Wanner(1996)}]{Hairer1996}
\bibinfo{author}{E.~Hairer}, \bibinfo{author}{G.~Wanner},
  \bibinfo{title}{Solving {O}rdinary {D}ifferential {E}quations {II}: {S}tiff
  and {D}ifferential-{A}lgebraic {P}roblems}, \bibinfo{publisher}{Springer},
  \bibinfo{year}{1996}.
\bibitem[{Jourdan et~al.(2014)Jourdan, Bencteux, and Adjanor}]{Jourdan2014}
\bibinfo{author}{T.~Jourdan}, \bibinfo{author}{G.~Bencteux},
  \bibinfo{author}{G.~Adjanor}, \bibinfo{journal}{J. Nucl. Mater.}
  \bibinfo{volume}{444} (\bibinfo{year}{2014}) \bibinfo{pages}{298}.
\bibitem[{Marian and Bulatov(2011)}]{Marian2011}
\bibinfo{author}{J.~Marian}, \bibinfo{author}{V.~V. Bulatov},
  \bibinfo{journal}{J. Nucl. Mater.} \bibinfo{volume}{415}
  (\bibinfo{year}{2011}) \bibinfo{pages}{84}.
\bibitem[{Gherardi et~al.(2012)Gherardi, Jourdan, Le~Bourdiec, and
  Bencteux}]{Gherardi2012}
\bibinfo{author}{M.~Gherardi}, \bibinfo{author}{T.~Jourdan},
  \bibinfo{author}{S.~Le~Bourdiec}, \bibinfo{author}{G.~Bencteux},
  \bibinfo{journal}{Comput. Phys. Commun.} \bibinfo{volume}{183}
  (\bibinfo{year}{2012}) \bibinfo{pages}{1966}.
\bibitem[{Dunn and Capolungo(2015)}]{Dunn2015}
\bibinfo{author}{A.~Y. Dunn}, \bibinfo{author}{L.~Capolungo},
  \bibinfo{journal}{Comp. Mater. Sci.} \bibinfo{volume}{102}
  (\bibinfo{year}{2015}) \bibinfo{pages}{314}.
\bibitem[{Kiritani(1973)}]{Kiritani1973}
\bibinfo{author}{M.~Kiritani}, \bibinfo{journal}{J. Phys. Soc. Jpn}
  \bibinfo{volume}{35} (\bibinfo{year}{1973}) \bibinfo{pages}{95}.
\bibitem[{Golubov et~al.(2001)Golubov, Ovcharenko, Barashev, and
  Singh}]{Golubov2001}
\bibinfo{author}{S.~I. Golubov}, \bibinfo{author}{A.~M. Ovcharenko},
  \bibinfo{author}{A.~V. Barashev}, \bibinfo{author}{B.~N. Singh},
  \bibinfo{journal}{Philos. Mag. A} \bibinfo{volume}{81} (\bibinfo{year}{2001})
  \bibinfo{pages}{643}.
\bibitem[{Ovcharenko et~al.(2003)Ovcharenko, Golubov, Woo, and
  Huang}]{Ovcharenko2003}
\bibinfo{author}{A.~Ovcharenko}, \bibinfo{author}{S.~Golubov},
  \bibinfo{author}{C.~Woo}, \bibinfo{author}{H.~Huang},
  \bibinfo{journal}{Comput. Phys. Commun.} \bibinfo{volume}{152}
  (\bibinfo{year}{2003}) \bibinfo{pages}{208 -- 226}.
\bibitem[{Golubov et~al.(2012)Golubov, Barashev, and Stoller}]{Golubov2012}
\bibinfo{author}{S.~I. Golubov}, \bibinfo{author}{A.~V. Barashev},
  \bibinfo{author}{R.~E. Stoller}, in: \bibinfo{editor}{R.~J. Konings} (Ed.),
  \bibinfo{booktitle}{Comprehensive Nuclear Materials},
  \bibinfo{publisher}{Elsevier}, \bibinfo{address}{Oxford},
  \bibinfo{year}{2012}, pp. \bibinfo{pages}{357 -- 391}.
\bibitem[{Waite(1957)}]{waite_theoretical_1957}
\bibinfo{author}{T.~R. Waite}, \bibinfo{journal}{Phys. Rev.}
  \bibinfo{volume}{107} (\bibinfo{year}{1957}) \bibinfo{pages}{463}.
\bibitem[{Russel(1971)}]{Russel1971}
\bibinfo{author}{K.~Russel}, \bibinfo{journal}{Acta Metallurgica}
  \bibinfo{volume}{19} (\bibinfo{year}{1971}) \bibinfo{pages}{753}.
\bibitem[{Goodrich(1964)}]{Goodrich1964}
\bibinfo{author}{F.~C. Goodrich}, \bibinfo{journal}{Proceedings of the Royal
  Society of London. Series A, mathematical and physical science}
  \bibinfo{volume}{277} (\bibinfo{year}{1964}) \bibinfo{pages}{167}.
\bibitem[{Terrier et~al.(????)Terrier, Ath\`enes, Jourdan, Adjanor, and
  Stoltz}]{Terrierxxxx}
\bibinfo{author}{P.~Terrier}, \bibinfo{author}{M.~Ath\`enes},
  \bibinfo{author}{T.~Jourdan}, \bibinfo{author}{G.~Adjanor},
  \bibinfo{author}{G.~Stoltz}, \bibinfo{journal}{in preparation}.
\bibitem[{Sharafat and Ghoniem(1990)}]{Sharafat1990}
\bibinfo{author}{S.~Sharafat}, \bibinfo{author}{N.~Ghoniem},
  \bibinfo{journal}{Radiat. Eff. Defects Solids} \bibinfo{volume}{113}
  (\bibinfo{year}{1990}) \bibinfo{pages}{331}.
\bibitem[{Ghoniem and Sharafat(1980)}]{Ghoniem1980}
\bibinfo{author}{N.~Ghoniem}, \bibinfo{author}{S.~Sharafat},
  \bibinfo{journal}{J. Nucl. Mater.} \bibinfo{volume}{92}
  (\bibinfo{year}{1980}) \bibinfo{pages}{121}.
\bibitem[{Ghoniem(1999)}]{Ghoniem1999}
\bibinfo{author}{N.~Ghoniem}, \bibinfo{journal}{Radiat. Eff. Defects Solids}
  \bibinfo{volume}{148} (\bibinfo{year}{1999}) \bibinfo{pages}{269}.
\bibitem[{Turkin and Bakai(2006)}]{Turkin2006}
\bibinfo{author}{A.~A. Turkin}, \bibinfo{author}{A.~S. Bakai},
  \bibinfo{journal}{J. Nucl. Mater.} \bibinfo{volume}{358}
  (\bibinfo{year}{2006}) \bibinfo{pages}{10}.
\bibitem[{Turkin and Bakai(2007)}]{Turkin2007}
\bibinfo{author}{A.~A. Turkin}, \bibinfo{author}{A.~S. Bakai},
  \bibinfo{journal}{Probl. At. Sci. Tech.} \bibinfo{volume}{3}
  (\bibinfo{year}{2007}) \bibinfo{pages}{394}.
\bibitem[{Fletcher(1991)}]{Fletcher1991}
\bibinfo{author}{C.~A.~J. Fletcher}, \bibinfo{title}{Computational Techniques
  for Fluid Dynamics 1: Fundamental and General Techniques},
  \bibinfo{publisher}{Springer-Verlag, Berlin}, \bibinfo{year}{1991}.
\bibitem[{Ozkan and Ortoleva(2000)}]{Ozkan2000}
\bibinfo{author}{G.~Ozkan}, \bibinfo{author}{P.~Ortoleva}, \bibinfo{journal}{J.
  Chem. Phys.} \bibinfo{volume}{112} (\bibinfo{year}{2000})
  \bibinfo{pages}{10510}.
\bibitem[{Chang and Cooper(1970)}]{Chang1970}
\bibinfo{author}{J.~Chang}, \bibinfo{author}{G.~Cooper}, \bibinfo{journal}{J.
  Comput. Phys.} \bibinfo{volume}{6} (\bibinfo{year}{1970}) \bibinfo{pages}{1}.
\bibitem[{Rempel et~al.(2009)Rempel, Bawendi, and Jensen}]{Rempel2009}
\bibinfo{author}{J.~Y. Rempel}, \bibinfo{author}{M.~G. Bawendi},
  \bibinfo{author}{K.~F. Jensen}, \bibinfo{journal}{J. Am. Chem. Soc.}
  \bibinfo{volume}{131} (\bibinfo{year}{2009}) \bibinfo{pages}{4479}.
\bibitem[{Johnson(1978)}]{Johnson1978}
\bibinfo{author}{R.~A. Johnson}, \bibinfo{journal}{J. Nucl. Mater.}
  \bibinfo{volume}{75} (\bibinfo{year}{1978}) \bibinfo{pages}{77}.
\bibitem[{Hindmarsh et~al.(2005)Hindmarsh, Brown, Grant, Lee, Serban, Shumaker,
  and Woodward}]{Hindmarsh2005}
\bibinfo{author}{A.~C. Hindmarsh}, \bibinfo{author}{P.~N. Brown},
  \bibinfo{author}{K.~E. Grant}, \bibinfo{author}{S.~L. Lee},
  \bibinfo{author}{R.~Serban}, \bibinfo{author}{D.~E. Shumaker},
  \bibinfo{author}{C.~S. Woodward}, \bibinfo{journal}{ACM Trans. Math. Softw.}
  \bibinfo{volume}{31} (\bibinfo{year}{2005}) \bibinfo{pages}{363--396}.
\bibitem[{Hairer and Wanner(1999)}]{Hairer1999}
\bibinfo{author}{E.~Hairer}, \bibinfo{author}{G.~Wanner}, \bibinfo{journal}{J.
  Comput. Appl. Math.} \bibinfo{volume}{111} (\bibinfo{year}{1999})
  \bibinfo{pages}{93}.
\bibitem[{Leveque(2004)}]{Leveque}
\bibinfo{author}{R.~J. Leveque}, \bibinfo{title}{{F}inite-{V}olume {M}ethods
  for {H}yperbolic {P}roblems}, \bibinfo{year}{2004}.
\bibitem[{Brailsford et~al.(1976)Brailsford, Bullough, and
  Hayns}]{Brailsford1976}
\bibinfo{author}{A.~D. Brailsford}, \bibinfo{author}{R.~Bullough},
  \bibinfo{author}{M.~R. Hayns}, \bibinfo{journal}{J. Nucl. Mater.}
  \bibinfo{volume}{60} (\bibinfo{year}{1976}) \bibinfo{pages}{246--256}.
\bibitem[{Kurganov et~al.(2001)Kurganov, Noelle, and Petrova}]{Kurganov2001}
\bibinfo{author}{A.~Kurganov}, \bibinfo{author}{S.~Noelle},
  \bibinfo{author}{G.~Petrova}, \bibinfo{journal}{SIAM J. Sci. Comput.}
  \bibinfo{volume}{23} (\bibinfo{year}{2001}) \bibinfo{pages}{707--740}.
\bibitem[{Kurganov and Tadmor(2000)}]{Kurganov2000}
\bibinfo{author}{A.~Kurganov}, \bibinfo{author}{E.~Tadmor},
  \bibinfo{journal}{J. Comput. Phys.} \bibinfo{volume}{160}
  (\bibinfo{year}{2000}) \bibinfo{pages}{241}.
\bibitem[{Suresh and Huynh(1997)}]{Suresh1997}
\bibinfo{author}{A.~Suresh}, \bibinfo{author}{H.~T. Huynh},
  \bibinfo{journal}{J. Comput. Phys.} \bibinfo{volume}{136}
  (\bibinfo{year}{1997}) \bibinfo{pages}{83--99}.
\bibitem[{Ardell(1988)}]{Ardell1988}
\bibinfo{author}{A.~J. Ardell}, in: \bibinfo{editor}{G.~W. Lorimer} (Ed.),
  \bibinfo{booktitle}{Phase transformations '87}.
\bibitem[{Lifshitz and Slyozov(1961)}]{Lifshitz1961}
\bibinfo{author}{I.~M. Lifshitz}, \bibinfo{author}{V.~V. Slyozov},
  \bibinfo{journal}{J. Phys. Chem. Solids} \bibinfo{volume}{19}
  (\bibinfo{year}{1961}) \bibinfo{pages}{35}.
\bibitem[{Wagner(1961)}]{Wagner1961}
\bibinfo{author}{C.~Wagner}, \bibinfo{journal}{Z. Elektrochem.}
  \bibinfo{volume}{65} (\bibinfo{year}{1961}) \bibinfo{pages}{581}.
\bibitem[{Kirchner(1973)}]{Kirchner1973}
\bibinfo{author}{H.~O.~K. Kirchner}, \bibinfo{journal}{Acta Metall.}
  \bibinfo{volume}{21} (\bibinfo{year}{1973}) \bibinfo{pages}{85}.
\bibitem[{Burton and Speight(1986)}]{Burton1986}
\bibinfo{author}{B.~Burton}, \bibinfo{author}{M.~V. Speight},
  \bibinfo{journal}{Philos. Mag. A} \bibinfo{volume}{53} (\bibinfo{year}{1986})
  \bibinfo{pages}{385}.
\bibitem[{Koiwa(1974)}]{Koiwa1974}
\bibinfo{author}{M.~Koiwa}, \bibinfo{journal}{J. Phys. Soc. Jpn}
  \bibinfo{volume}{37} (\bibinfo{year}{1974}) \bibinfo{pages}{1532}.
\bibitem[{Hirth and Lothe(1968)}]{Hirth1968}
\bibinfo{author}{J.~P. Hirth}, \bibinfo{author}{J.~Lothe},
  \bibinfo{title}{Theory of dislocations}, \bibinfo{publisher}{McGraw-Hill},
  \bibinfo{year}{1968}.
\bibitem[{Seeger and G\"osele(1977)}]{Seeger1977}
\bibinfo{author}{A.~Seeger}, \bibinfo{author}{U.~G\"osele},
  \bibinfo{journal}{Phys. Lett. A} \bibinfo{volume}{61} (\bibinfo{year}{1977})
  \bibinfo{pages}{423}.
\bibitem[{Dubinko et~al.(2005)Dubinko, Abysov, and Turkin}]{Dubinko2005}
\bibinfo{author}{V.~I. Dubinko}, \bibinfo{author}{A.~S. Abysov},
  \bibinfo{author}{A.~A. Turkin}, \bibinfo{journal}{J. Nucl. Mater.}
  \bibinfo{volume}{336} (\bibinfo{year}{2005}) \bibinfo{pages}{11}.
\bibitem[{Jourdan(2015)}]{Jourdan2015}
\bibinfo{author}{T.~Jourdan}, \bibinfo{journal}{J. Nucl. Mater.}
  \bibinfo{volume}{467} (\bibinfo{year}{2015}) \bibinfo{pages}{286}.
\bibitem[{Ehrhart et~al.(1991)Ehrhart, Jung, Schultz, and
  Ullmaier}]{Ehrhart1991}
\bibinfo{author}{P.~Ehrhart}, \bibinfo{author}{P.~Jung},
  \bibinfo{author}{H.~Schultz}, \bibinfo{author}{H.~Ullmaier},
  \bibinfo{title}{Landolt--B{\"o}rnstein, Numerical Data and Functional
  Relationships in Science and Technology}, Atomic Defects In Metals,
  \bibinfo{publisher}{Springer}, \bibinfo{year}{1991}.
\bibitem[{Bak\'o et~al.(2011)Bak\'o, Clouet, Dupuy, and Bl\'etry}]{Bako2011}
\bibinfo{author}{B.~Bak\'o}, \bibinfo{author}{E.~Clouet},
  \bibinfo{author}{L.~M. Dupuy}, \bibinfo{author}{M.~Bl\'etry},
  \bibinfo{journal}{Philos. Mag.} \bibinfo{volume}{91} (\bibinfo{year}{2011})
  \bibinfo{pages}{3173}.
\bibitem[{Berthier et~al.(2011)Berthier, Maras, Braems, and
  Legrand}]{Berthier2011}
\bibinfo{author}{F.~Berthier}, \bibinfo{author}{E.~Maras},
  \bibinfo{author}{I.~Braems}, \bibinfo{author}{B.~Legrand},
  \bibinfo{journal}{Solid State Phenom.} \bibinfo{volume}{172-174}
  (\bibinfo{year}{2011}) \bibinfo{pages}{664}.
\bibitem[{Senkov(2008)}]{Senkov2008}
\bibinfo{author}{O.~N. Senkov}, \bibinfo{journal}{Scr. Mater.}
  \bibinfo{volume}{59} (\bibinfo{year}{2008}) \bibinfo{pages}{171}.
\bibitem[{Bonafos et~al.(1998)Bonafos, Mathiot, and Claverie}]{Bonafos1998}
\bibinfo{author}{C.~Bonafos}, \bibinfo{author}{D.~Mathiot},
  \bibinfo{author}{A.~Claverie}, \bibinfo{journal}{J. Appl. Phys.}
  \bibinfo{volume}{83} (\bibinfo{year}{1998}) \bibinfo{pages}{3008}.
\bibitem[{Pan et~al.(1997)Pan, Tu, and Prussin}]{Pan1997}
\bibinfo{author}{G.~Z. Pan}, \bibinfo{author}{K.~N. Tu},
  \bibinfo{author}{A.~Prussin}, \bibinfo{journal}{J. Appl. Phys.}
  \bibinfo{volume}{81} (\bibinfo{year}{1997}) \bibinfo{pages}{78}.
\bibitem[{Brailsford and Wynblatt(1979)}]{brailsford_dependence_1979}
\bibinfo{author}{A.~D. Brailsford}, \bibinfo{author}{P.~Wynblatt},
  \bibinfo{journal}{Acta Metall} \bibinfo{volume}{27} (\bibinfo{year}{1979})
  \bibinfo{pages}{489}.
\bibitem[{Golubov et~al.(2007)Golubov, Stoller, Zinkle, and
  Ovcharenko}]{Golubov2007}
\bibinfo{author}{S.~Golubov}, \bibinfo{author}{R.~Stoller},
  \bibinfo{author}{S.~Zinkle}, \bibinfo{author}{A.~Ovcharenko},
  \bibinfo{journal}{J. Nucl. Mater.} \bibinfo{volume}{261}
  (\bibinfo{year}{2007}) \bibinfo{pages}{149}.
\bibitem[{Capdeville(2008)}]{Capdeville2008}
\bibinfo{author}{G.~Capdeville}, \bibinfo{journal}{J. Comput. Phys.}
  \bibinfo{volume}{227} (\bibinfo{year}{2008}) \bibinfo{pages}{2977--3014}.

\end{thebibliography}

\end{document}